# Orbital Angular Momentum in the Nucleon

Gerald T. Garvey
Los Alamos National Laboratory
*Los Alamos, New Mexico, 87545*

Analysis of the measured value of the integrated $\bar{d} - \bar{u}$ asymmetry ($I_{fas}$ = 0.147±0.027) in the nucleon show it to arise from nucleon fluctuations into baryon plus pion. Requiring angular momentum conservation in these fluctuations shows the associated orbital angular momentum is equal to the value of the flavor asymmetry.

The partonic composition of nucleon spin has continued to occupy the attention of physicists for the past 20 years[1]. The measured asymmetry in deep inelastic lepton scattering off polarized nucleonic targets convincingly demonstrates that the summed projection of quark spins is appreciably less than the projection of the nucleon angular momentum. This short fall was in earlier times termed a "spin crisis". Currently the value of projected spin of the quarks on the total angular momentum of the proton is taken to be $\Delta\Sigma/2 = (\Delta u + \Delta d + \Delta s)/2 = 0.183 \pm 0.0085$[2]. Thus the projected spin carried by quarks is observed to be ~ 40% of the proton's total angular momentum. Theory and experiment continue to investigate where the rest of the proton's angular momentum might reside. The most recent data suggests that the spin carried by gluons ($\Delta g$) is small [2] so one must look to orbital angular momentum (OAM) carried by quarks and gluons to account for the smaller than expected angular momentum found on quark spin.

The proton's total AM can be written as

$$\vec{J} = \vec{S}_q + \vec{L}_q + \vec{S}_g + \vec{L}_g \quad (1)$$

with corresponding projections along the total AM

$$J_3 = \frac{\Delta\Sigma}{2} + L_{q,3} + S_{g,3} + L_{g,3} \quad (2)$$

$L_{q,3}$ is the projection of quark OAM, with the latter two terms being the projection of spin and OAM of gluons. At the present time there exists no measurement of the OAM carried by quarks or gluons. Measuring the OAM is experimentally difficult [3], but may be possible by measuring generalized parton distributions [4]. Calculation of the quark OAM has recently been carried out on the lattice [5]. In the following it will be shown shown that the OAM created in hadronic fluctuations into baryon + Goldstone boson configurations can be an important ingredient in confronting the smaller than expected value of the spin carried by quarks.

The data fixing the quark spin contributions has three sources; semi-leptonic axial weak decays determine the following spin combinations: 1) SU(2), ($\Delta\Sigma_{SU(2)} = \Delta u - \Delta d$) and 2) SU(3) ($\Delta\Sigma_{SU(3)} = \Delta u + \Delta d - 2\Delta s$), and 3) the asymmetry in DIS from longitudinally polarized nucleons. The approximately $Q^2 = 0$, axial weak decay rates of the nucleon and hyperons contain all the correlations present in those baryonic systems while the result extracted from DIS directly accesses the charge weighted quark spins essentially free of correlations. The approach employed below invokes a model with specific correlations (pions) that quantitatively accounts for the observed properties of the $\bar{u}, \bar{d}$ difference in the proton to obtain a contribution to the OAM in the nucleon.

Measurement of the flavor asymmetry in the proton via muon DIS [5] showed a large violation of the Gottfried sum rule.

$$I_G(0,1:Q^2) = \int_0^1 \frac{F_2^p(x,Q^2) - F_2^n(x,Q^2)}{x} dx$$
$$= \frac{1}{3} - \frac{2}{3} \int_0^1 [\bar{d}(x,Q^2) - \bar{u}(x,Q^2)] dx = \frac{1}{3} - \frac{2}{3} I_{fas} \quad (3)$$

The measured value [6] for the integral over the sea quarks is $I_{fas} = 0.176 \pm 0.038$. A Drell-Yan experiment [7], that directly accessed this flavor asymmetry in the sea determined $I_{fas} = 0.118 \pm 0.038$. The weighted average of the two results for the integrated flavor asymmetry is $I_{fas} = 0.147 \pm 0.027$. This large value of the asymmetry came as a surprise to many who believed that the quark sea evolved from gluon pair production, creating a sea that should be very nearly symmetric. The observed asymmetry in the sea is sometimes incorrectly referred as an isospin violation; it is rather a consequence of isospin conservation. It was pointed out much earlier [8] that pionic fluctuations would create such a asymmetry in the sea and was subsequently calculated [9] before the experiments referenced above were carried out. Further, and of importance for what follows, the appearance of the flavor asymmetry at large values of x (0.1< x <0.25) indicates that the pions are fluctuated off the whole nucleon rather than off individual valence quarks

[10,11,12] as would be the case in some versions of chiral perturbation theory ($\chi$PT) [13]. While pions constitute a small fraction of the total sea, they have important consequences such as the asymmetry discussed above. The flavor asymmetry cannot readily be produced via evolution from valence quarks. The asymmetry must be put into parton distributions "by hand" [14,15,16], further, as the flavor asymmetry is non-singlet its value is conserved when integrated over all x. As shown below accepting the pionic basis for the sea quark asymmetry has consequences for the spin carried by quarks. An extensive discussion of theoretical approaches calculating the flavor asymmetry can be found in review articles [17,18].

Extending the simple constituent quark model to include pionic fluctuations of the proton into N-$\pi$ and $\Delta$–$\pi$ configurations leads to

$$|p\rangle = \frac{1}{\sqrt{1+a^2+b^2}}\{|p_0\rangle + a(-\sqrt{\frac{1}{3}}|p_0\pi^0\rangle + \sqrt{\frac{2}{3}}|n_0\pi^+\rangle)$$
$$+ b(\sqrt{\frac{1}{2}}|\Delta_0^{++}\pi^-\rangle - \sqrt{\frac{1}{3}}|\Delta_0^+\pi^0\rangle + \sqrt{\frac{1}{6}}|\Delta_0^0\pi^+\rangle)\} \quad (4)$$

Where the 0 subscript on the baryons indicates that they are to be taken as the constituent quark only configurations. The amount of fluctuation of the proton into N$\pi$ is $a^2$, while $b^2$ is the amount into $\Delta\pi$. The amplitudes of the various pionic charge states are the vector addition coefficients that conserve isospin, In this model it is easy to show

$$I_{fas} = \int_0^1 [\bar{d}(x) - \bar{u}(x)]dx = \frac{2a^2 - b^2}{3(1+a^2+b^2)} \quad (5)$$

Values calculated for $a^2$ and $b^2$ in standard prescriptions [9,10] are not inconsistent with the experimental values cited above.

The presence of such pionic fluctuations introduces orbital angular momentum into the nucleon. The conservation of angular momentum and parity, requires the relative angular momentum of the pion-baryon be l=1. The projection of this OAM on the total AM is fixed by angular momentum conservation and is,

$$\langle p\uparrow|l_3|p\uparrow\rangle = \frac{2a^2 - b^2}{3(1+a^2+b^2)} \quad (6)$$

This value is identical with that obtained for the integral ($I_{fas}$) of the asymmetry, because the spin and isospin of the nucleon are both 1/2, the spin and isospin of the delta are both 3/2 and the value of the OAM is the same as the pion isospin. Thus the orbital angular momentum projection associated with pionic fluctuations of the nucleon is appreciable (0.147±0.027). Such OAM is presumably increased by $\Lambda k$, and $\Sigma k$ fluctuations. These produce no contribution to the flavor asymmetry, but as the kaon is $J^\pi$= 0$^-$ and the $J^\pi$ of $\Lambda$ and $\Sigma$ is 1/2$^+$, they serve to increase the OAM . Total angular momentum conservation requires that the OAM projection is opposite the spin projection so that 0.147±0.027 of the missing quark spin appears as OAM. Apparently contradictory to this finding, recent lattice gauge calculations (LGC) [5] find the angular momentum carried by up and down quarks to be individually large (|~0.15|) , approximately equal but of opposite sign so the resulting OAM carried by u and d quarks is very small (~0). The LGC employed pions and their result appears independent of pion mass so the difference with our analysis is interesting. In our model it is not clear as how to apportion the OAM among the partons as it is in the relative motion of the pion and baryon.

Altering the spin of u and d quarks also affects the nucleon's axial coupling constant ($G_A$). The effect is given below, where a large contribution comes from the interference between the N$\pi$ and $\Delta\pi$ Fock components;

$$G_A = \langle p\uparrow|\sigma_3\tau_3|p\uparrow\rangle = \Delta u - \Delta d$$
$$= \frac{1}{1+a^2+b^2}(\frac{5}{3} + \frac{5a^2}{27} + \frac{25b^2}{27} + \frac{32ab\sqrt{2}}{27}) \quad (8)$$

The calculation is carried out incorporating spin and orbital angular momentum into eq 4, conserving AM. To determine $G_A$, values for $a$ and $b$ must be individually specified. Figure 1 shows the values of $a$ extracted from the flavor asymmetry as a function of $n$ where $b^2=a^2/n$. It appears that values of $a^2$ are unrealistically large ($a^2>.4$) for certain values of the flavor asymmetry and the relative fraction of N$\pi$/$\Delta\pi$ to be compatible with the observations in DIS (19).

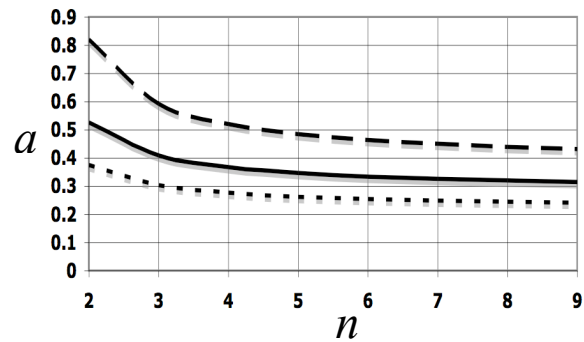

Figure 1 Shows the amount of $Np$ configuration ($a^2$) in the proton as a function of $n=a^2/b^2$ using the measured value of the flavor asymmetry. The middle curve is for the central value of the measured asymmetry with upper and lower curves corresponding to the uncertainty in the value of the flavor asymmetry.

Figure 2 shows that the effects of the pionic fluctuations are rather stable in their impact on $G_A$. It is evident that while providing some reduction from $G_A=5/3$ the effects of the pionic fluctuation do not account for the observed value of 1.267. We believe this to be appropriate as there are other effects that come into play to further reduce $G_A$ [20,21].

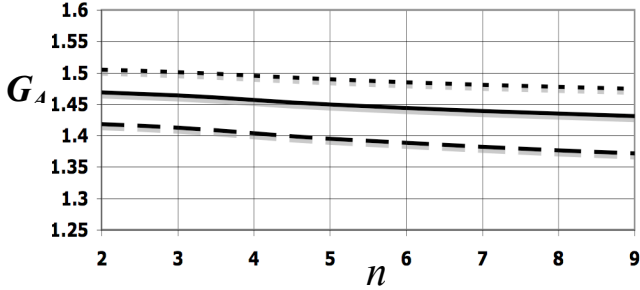

**Figure 2.** Shows the effect of the flavor asymmetry and the fraction of $N\pi/\Delta\pi$ on the axial coupling constant $G_A$. The upper curve corresponds to $I_{fas}=0.120$, the middle to $0.146$ and the lower to $0.174$

One could be concerned that a large value for pionic fluctuations with its consequences for both spins and orbital angular momentum would seriously affect the value obtained for $\mu_p/\mu_n$. It is measured to be -1.46., The simple SU(2) constituent quark model gives -1.50. In the model used here one finds for the ratio,

$$-\frac{\mu_p}{\mu_n} = \frac{3}{2}\left(\frac{1+\frac{13}{27}a^2 + \frac{23}{27}b^2 + \frac{16}{27}ab\sqrt{2}}{1+\frac{8}{9}a^2 + \frac{4}{9}b^2 + \frac{16}{27}ab\sqrt{2}}\right) \quad (9)$$

In deriving eq.(9) it is assumed that the effective virtual pion mass ~equals $m_p/3$. The effect of adding the pionic fluctuations is shown in Fig.3, where it is evident that the pionic fluctuations as modeled in this paper are as close to experiment as the simpler model.

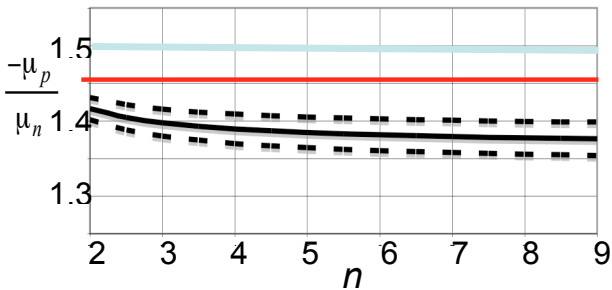

**Figure 3.** The value of $-\mu_p/\mu_n$ as a function of $n$ where $b^2=a^2/n$. The upper dashed curve corresponds to a asymmetry of $I_{fas}=0.12$, with the two below corresponding to $I_{fas}= 0.146$ and $0.176$ respectively. The red line indicates the measured value of $\mu_p/\mu_n$ and the blue, the value of the constituent model

A least squares fit [22] to baryon magnetic moments and axial transition rates in which pions were added to SU(2) baryonic wave functions generated results similar to the results presented here. Including large pionic fluctuations and a numerical equality for the value of the pionic orbital angular momentum and the $\bar{d}-\bar{u}$ asymmetry.

The assumption (fact) that the pion is emitted by the nucleon rather than off individual constituent quarks has significant consequences. Eichten et. al [13] addressed the flavor asymmetry in the context of a model of $\chi$PT, where the degrees of freedom are constituent quarks and pions. In $\chi$PT, the flavor asymmetry is generated by the fluctuation of individual constituent quarks into quark + pion. Respecting isospin, this model directly produces the requirement that $\bar{d}/\bar{u}<11/7$. Fig. 3 of ref. [6] shows that this condition would require a negligible contribution from the symmetric sea for $0.15 < x < 0.25$, which is likely not the case. The reason for this upper limit on $\bar{d}/\bar{u}$ is that this approach requires too large a contribution from baryonic T=3/2 states ($a^2/b^2=1.25$), which suppress the asymmetry. As a result this approach has the flavor integrated flavor asymmetry of the proton given by

$$I_{fas} = \frac{2}{3}\alpha$$

where the probability of a constituent quark fluctuating into a pion is $3\alpha/2$. With $\alpha=0.225\pm0.040$ the probability that the proton remains as 3 bare constituent quarks is just $(1-9\alpha/2)$ a small and unlikely result. Reference [13] further noted that the existence of a flavor asymmetry has consequences for the spin carried by quarks and investigated this within the context of $\chi$PT. In $\chi$PT the quarks have a definite helicity which is flipped upon the emission of a Goldstone boson so that

$$\Delta u_{\chi PT} = \frac{4}{3} - \frac{7\alpha}{3}$$

$$\Delta d_{\chi PT} = \frac{-1}{3} - \frac{2\alpha}{3}$$

so $\Delta u_{\chi PT}+\Delta d_{\chi PT}=1-3\alpha= 0.325\pm0.120$, not out of line with the observed spin quenching. This value will be further reduced by the emission of other Goldstone bosons not contributing to the flavor asymmetry. This approach also yields $G_A=\Delta u_{\chi PT}-\Delta d_{\chi PT}=5/3-5\alpha/3=1.29\pm0.045$, again fairly close to experiment. Extensions of this approach were carried out by a variety of authors [23,24,25] but none explicitly address the issue of OAM. If we ascribe OAM to the pion-baryon relative motion in order to preserve the spin of the initial state then

$$\Delta l_{\chi PT} = \frac{3\alpha}{2} = +.34 \pm .06$$

In a somewhat different approach, Qing et al [26] extended the simple constituent quark model to include $q\bar{q}$ configurations. They further restricted these $q\bar{q}$ configurations to color singlet pseudo scalars, while the $qqq$ are color singlets within the octet and decoplet spaces. They fit the parameters of their dynamical model to the octet and decoplet ground state properties. They find a value for the OAM of $L_{q,3}$=0.047. a value appreciably smaller than the analysis above.

In conclusion, we have shown that accepting the measured asymmetry in the sea is due to pions and assuming these pions are emitted off the nucleon, then isospin and angular momentum conservation directly fix a value for the OAM associated with the asymmetry. OAM reduces $\Sigma$ from 1 to 0.71, which when coupled with other effects [20,21] may well account for the small value observed for $\Sigma$ The effect of these pionic fluctuations on $G_A$ and $\mu_p/\mu_n$ is reasonable and acceptable.

The author wishes to acknowledge discussions with Bogdan Povh that initiated this work. The support of the Medium Energy Program in the Nuclear Physics Division of the Office Science/DOE is also gratefully acknowledged.